\documentclass[12pt,preprint,natbib,iop,numberedappendix,appendixfloats]{emulateapj} 
\usepackage{graphicx,epsfig,amsmath} 

\lefthead{van~de~Ven \& van~der~Wel}
\righthead{}
\begin{document}

\newcommand{\gala}{{\tt GALAPAGOS}~}
\newcommand{\gf}{{\tt GALFIT}~}
\newcommand{\mh}{H_{\rm{F160W}}}
\newcommand{\reff}{R_{\rm{eff}}}
\newcommand{\msol}{M_{\odot}}
\newcommand{\msola}{10^{11}~M_{\odot}}
\newcommand{\msolb}{10^{10}~M_{\odot}}
\newcommand{\msolc}{10^{9}~M_{\odot}}
\newcommand{\ang}{\rm{\AA}}

\newcommand{\masyr}{mas\,yr$^{-1}$}
\newcommand{\kms}{km\,s$^{-1}$}
\newcommand{\dgr}{$^\circ$}
\newcommand{\Msun}{M$_\odot$}
\newcommand{\Lsun}{L$_\odot$}
\newcommand{\Lsunpcsq}{L$_\odot$\,pc$^{-2}$}
\newcommand{\Msunpcsq}{M$_\odot$\,pc$^{-2}$}
\newcommand{\Lsunpccube}{L$_\odot$\,pc$^{-3}$}
\newcommand{\Msunpccube}{M$_\odot$\,pc$^{-3}$}
\newcommand{\MLsun}{M$_\odot$/L$_\odot$}
\newcommand{\hkpc}{$h^{-1}$\,kpc}
\newcommand{\hMpc}{$h^{-1}$\,Mpc}
\newcommand{\hMsun}{$h^{-1}$\,M$_\odot$}

\newcommand{\rmd}{\mathrm{d}}
\newcommand{\rmed}{r_\mathrm{med}}
\newcommand{\eps}{\varepsilon}
\newcommand{\sphq}{\sin^2\varphi}
\newcommand{\cphq}{\cos^2\varphi}
\newcommand{\sthq}{\sin^2\vartheta}
\newcommand{\cthq}{\cos^2\vartheta}

\newcommand{\hst}{\texttt{HST}}
\newcommand{\sauron}{\texttt{SAURON}}
\newcommand{\atlas}{\texttt{ATLAS$^{3D}$}}
\newcommand{\ppak}{\texttt{PPAK}}
\newcommand{\califa}{\texttt{CALIFA}}
\newcommand{\sdss}{\texttt{SDSS}}
\newcommand{\sfourg}{\texttt{S$^4$G}}


\title{Deprojecting Sersic Profiles for Arbitrary Triaxial Shapes: Robust Measures of Intrinsic and Projected Galaxy Sizes}

\author{Glenn van~de~Ven\altaffilmark{1,3}}
\author{Arjen van~der~Wel\altaffilmark{2,3}}

\altaffiltext{1}{Department of Astrophysics, University of Vienna, T\"urkenschanzstrasse 17, 1180 Vienna, Austria; glenn.vandeven@univie.ac.at}
\altaffiltext{2}{Sterrenkundig Observatorium, Department of Physics and Astronomy, Ghent University, Belgium; arjen.vanderwel@ugent.be}
\altaffiltext{3}{Max-Planck Institut f\"ur Astronomie, K\"onigstuhl
  17, D-69117, Heidelberg, Germany}

\begin{abstract}
We present the analytical framework for converting projected light distributions with a S\'ersic profile into three-dimensional light distributions for stellar systems of arbitrary triaxial shape. The main practical result is the definition of a simple yet robust measure of intrinsic galaxy size: the median radius $\rmed$, defined as the radius of a sphere that contains 50\% of the total luminosity or mass, that is, the median distance of a star to the galaxy center. We examine how $\rmed$ depends on projected size measurements as a function of S\'ersic index and intrinsic axis ratios, and demonstrate its relative independence of these parameters. As an application we show that the projected semi-major axis length of the ellipse enclosing 50\% of the light is an unbiased proxy for $\rmed$, with small galaxy-to-galaxy scatter of $\sim$10\% (1$\sigma$), under the condition that the variation in triaxiality within the population is small. For galaxy populations with unknown or a large range in triaxiality an unbiased proxy for $\rmed$ is $1.3\times R_{e}$, where $R_{e}$ is the circularized half-light radius, with galaxy-to-galaxy scatter of 20-30\% (1$\sigma$). We also describe how inclinations can be estimated for individual galaxies based on the measured projected shape and prior knowledge of the intrinsic shape distribution of the corresponding galaxy population. We make the numerical implementation of our calculations available.
\end{abstract}

\section{Introduction}
\label{S:intro}

The spatial distribution of stars in a galaxy encodes key information
about its formation history, whether dissipative or dissipationless
processes dominated and whether angular momentum has been retained or
lost.  The half-light radius is the simplest observational measure of stellar mass
distribution. From an empirical perspective, this quantity has played
a central role in defining the nature of galaxies through examining
and interpreting their scaling relations \citep[e.g.,][]{kormendy77,
  djorgovski87, dressler87} and tracking the build up of galaxies
through cosmic time \citep[e.g.,][]{trujillo04, van-der-wel14}. From a
theoretical perspective, galaxy sizes have repeatedly revealed that
essential physical elements are missing in galaxy formation models
\citep[e.g.,][]{navarro00}, leading to the implementation of 
(astro)physical processes such as feedback.

Measurements of galaxy sizes as traced by stellar light have now
reached a point where 0.1 dex accurate statements about two-dimensional (2D) projected galaxy sizes across cosmic time can be made with great precision \cite[e.g.,][]{Mowla2019,van-der-wel14}.
Currently, we are limited by the interpretation of the data, not its
scarcity -- almost a unicum in the field of galaxy evolution.  One
limiting factor is the missing link between the projected size (the
half-light radius that we measure) and a physically more directly
meaningful quantity such as the average or median distance of a star
to the center of its galaxy.  This conversion factor, from projected
to three-dimensinonal (3D) intrinsic size, can vary by large factors given the range of possible galaxy shapes and projections. As a result, comparisons with simulations have remained indirect and prone to systematic effects \cite[e.g.,][]{Parsotan2021,Ludlow2019,Genel2018}.

We note that we take the word \emph{size} to mean the 2D projected or 3D intrinsic half-light radius (or, ideally, half-mass radius) that serves to quantify what we define as the mean radius of the galaxy. Other definitions of size adopt a fixed surface brightness or density threshold, aiming to quantify the \emph{full extent} of the galaxy, which is a useful but qualitatively different quantity from the mean radius.

To our knowledge attempts at the deprojection of measured 2D light
profiles so far have assumed spherical symmetry
\citep[e,g,][]{bezanson09}, symplifying the calculation, but failing
to account for the fact that in reality most galaxies are far from
spherical: even most early-type galaxies have an intrinsic
short-to-long axis ratio of no more than $\sim 0.3$
\citep[e.g.,][]{vincent05}, which more recently has been shown to be
the case across cosmic time \citep[e.g.,][]{chang13a}.  In this paper
we present the analytical framework and numerical implementation for
the conversion of 2D light profiles to 3D light distributions for
galaxies of arbitrary triaxial shape. Only with this machinery can
we take full advantage of the available high-quality data and make
accurate comparisons with theoretical predictions.

This papers is organised as follows: Section 2 describes how triaxial shapes project in 2D; Section 3 depicts the deprojection of S\'ersic profiles; Section 4
introduces the definition and derivation of the median radius,
$\rmed$; Section 5 outlines how to infer, in practice,
$\rmed$ from projected size and shape measuremts; Section 6 concludes with a brief description of implications of our findings for galaxy size estimates.  

\section{Ellipsoidal Shapes}
\label{S:ellipsoidal-shapes}

We consider intrinsic density distributions $\rho(x,y,z) = \rho(m)$ that are
constant on ellipsoids
\begin{equation}
  \label{eq:ellipsoid}
  m^2 = \frac{x^2}{a^2} + \frac{y^2}{b^2} + \frac{z^2}{c^2},
\end{equation}
with $a \ge b \ge c$. The major semi-axis length $a$ is a scale
parameter, whereas the intermediate-over-major ($p \equiv b/a$) and
minor-over-major ($q \equiv c/a$) axis ratios determine the intrinsic shape. In the
oblate or prolate axisymmetric limit we have $a=b>c$ (pancake-shaped)
or $a>b=c$ (cigar-shaped), respectively, while in the spherical
limit $a=b=c$ (so then $m=r/a$).  Note that $m$ is a dimensionless
ellipsoidal radius. 

We introduce a new Cartesian coordinate system $(x'',y'',z'')$, with
$x''$ and $y''$ in the plane of the sky and the $z''$-axis along the
line-of-sight. Choosing the $x''$-axis in the $(x,y)$-plane of the
intrinsic coordinate system \citep[cf.][]{de-zeeuw89} and their
Fig.~2), the transformation between both coordinate systems is known
once two viewing angles, the polar angle $\vartheta$ and azimuthal
angle $\varphi$, are specified.  The intrinsic $z$-axis projects onto
the $y''$-axis; for an axisymmetric galaxy model the $y''$-axis aligns
with the short axis of the projected density,\footnote{For an oblate
  galaxy, the alignment ($\psi=0$) follows directly from
  equation~\eqref{eq:misalignment_psi} since $T=0$. For a prolate
  galaxy it is most easily seen by exchanging $a$ and $c$ ($c>b=a$),
  so that the $z$-axis is again the symmetry axis (instead of the
  $x$-axis).}  but for a triaxial galaxy model the $y''$-axis is
misaligned by an angle $\psi \in [-\pi/2,\pi/2]$ such that 
\citep[cf.\ equation~B9 of][]{franx88}
\begin{equation}
  \label{eq:misalignment_psi}
    \tan2\psi = \frac{T\sin2\varphi\cos\vartheta}
    {\sthq + T\left(\sphq\cthq-\cphq\right)}\ ,
\end{equation}
where $T$ is the triaxiality parameter defined as $T =
(a^2-b^2)/(a^2-c^2)$. A rotation through $\psi$ transforms the
coordinate system $(x'',y'',z'')$ to $(x',y',z')$ such that the $x'$
and $y'$ axes are aligned with the major and minor axes of the
projected density (respectively), while $z'=z''$ is along the
line-of-sight.

Projecting $\rho(m)$ along the line-of-sight yields a surface
density $\Sigma(x',y') = \Sigma(m')$ that is constant on ellipses
in the sky-plane,
\begin{equation}
  \label{eq:surfmassell}
  \Sigma(m')  
  = \int_{-\infty}^{\infty} \rho(m) \, \rmd z'
  = \frac{abc}{a'b'} \; 2 \int_{0}^{\infty} 
  \rho(m) \, m \; \rmd u,
\end{equation}
where we have used $z' = abc \, \sinh(u)/(a'b') + \mathrm{constant}$,
and $m = m' \cosh(u)$. The sky-plane ellipse is given by
\begin{equation}
  \label{eq:ellipse}
  m'^2 = \frac{x'^2}{a'^2} + \frac{y'^2}{b'^2}.
\end{equation}
The projected major and minor semi-axis lengths, $a'$ and $b'$,
depend on the intrinsic semi-axis lengths $a$, $b$, and $c$ and
the viewing angles $\vartheta$ and $\varphi$ as follows:
\begin{equation}
  \label{eq:apbpellipse}
  a'^2 = \frac{2\,A^2}{B - \sqrt{B^2 - 4 A^2}},
  \qquad
  b'^2 = \frac{2\,A^2}{B + \sqrt{B^2 - 4 A^2}},
\end{equation}
where $A$ and $B$ are defined as
\begin{eqnarray}
  \label{eq:defA}
  A^2 & = & 
  a^2 b^2 \cthq + ( a^2 \sphq + b^2 \cphq ) c^2 \sthq,
  \\ \label{eq:defB}
  B & = &  
  a^2 (\cphq\cthq+\sphq)
  \nonumber \\  
  && + b^2 (\sphq\cthq+\cphq) + c^2 \sthq.
\end{eqnarray}
It follows that $A=a'b'$ is proportional to the area of the ellipse.

The flattening $q' \equiv b'/a'$ of the projected ellipses actually depends
on the viewing angles $(\vartheta,\varphi)$ and the intrinsic axis
ratios $(b/a,c/a)$, and is independent of the scale length.  Since
the intrinsic and projected semi-major axis lengths are directly
related via the left equation of \eqref{eq:apbpellipse}, the scale
length can be set by choosing either $a$ or $a'$.

In what follows, we will describe the density and related quantities in terms of mass, including intrinsic mass density, surface mass density and enclosed mass. Without loss of generality, these quantities may also be expressed in terms of light, i.e., intrinsic luminosity density, surface brightness, and enclosed luminosity. However, mixing of mass and light quantities requires an additional conversion with the mass-to-light ratio which in general varies with galactic radius.

\begin{figure}[t]
\begin{center}
 \includegraphics[width=0.48\textwidth]{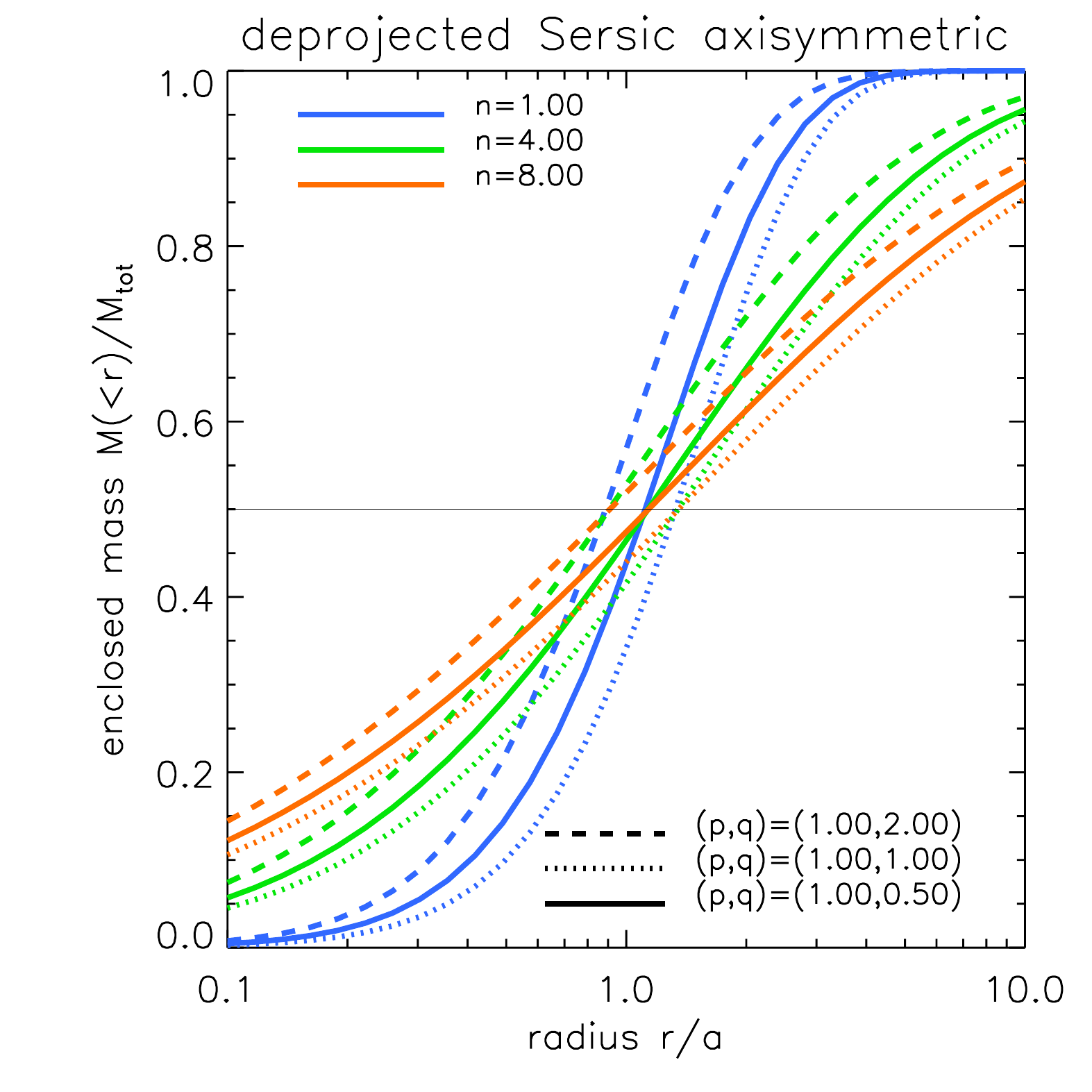}
 \includegraphics[width=0.48\textwidth]{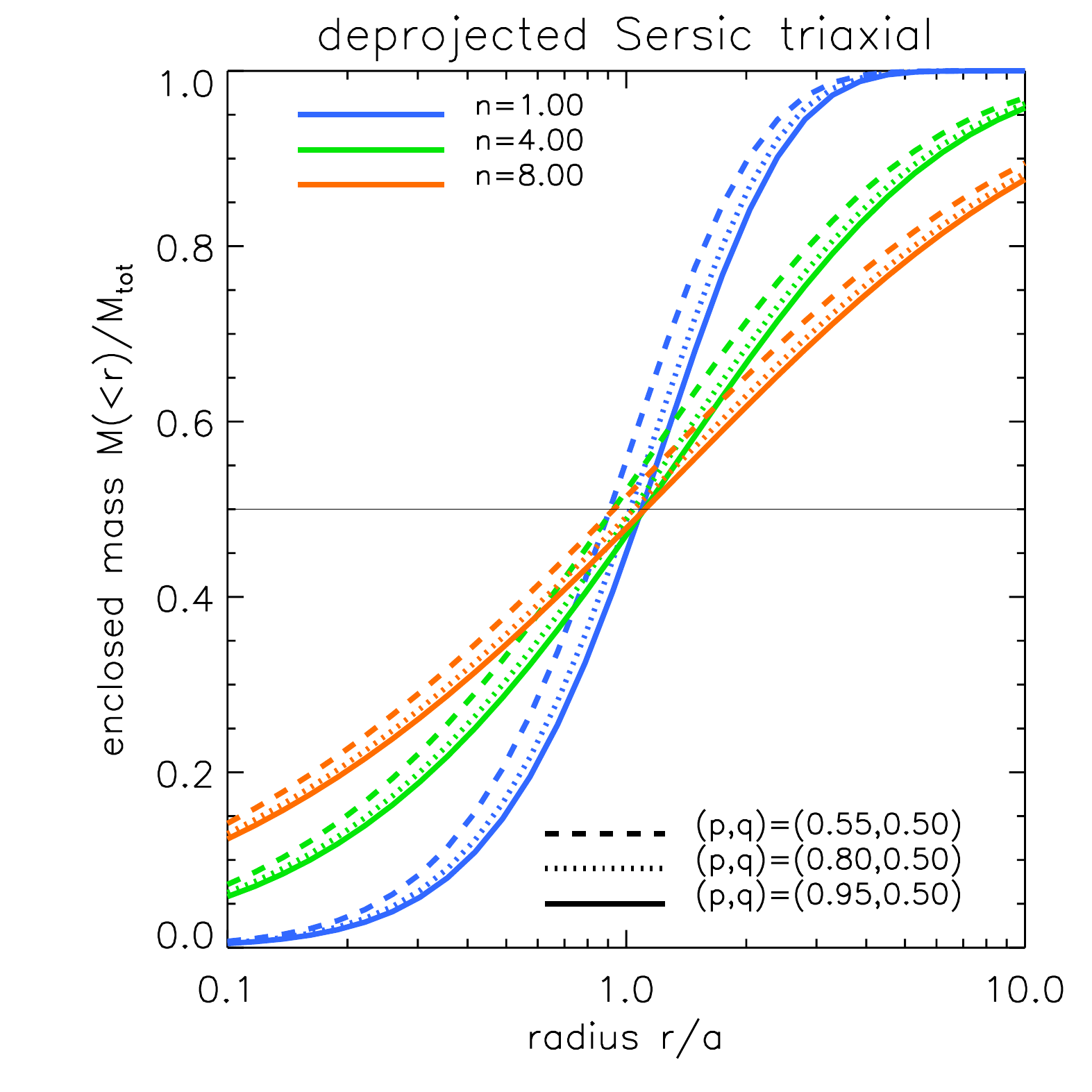}
  \caption{\small De-projected S\'ersic profile. The enclosed
    mass $M(<r)$ normalized by the total mass
    $M_\mathrm{tot}$, within a spherical radius $r$ normalized by the
    scale length $a$. As indicated, the colors represent different
    S\'ersic index $n$ and the line styles different intrinsic shapes
    given by the intrinsic axis ratios $p=b/a$ and $q=c/a$. The
    \emph{top} panel is for spheroidal shapes, either oblate axisymmetric($q<1$)
    or prolate axisymmetric ($q>1$), and the \emph{bottom} panel for fully triaxial
    shapes. }
  \label{fig:masssph}
\end{center}
\end{figure}

\begin{figure}[t]
\begin{center}
 \includegraphics[width=0.48\textwidth]{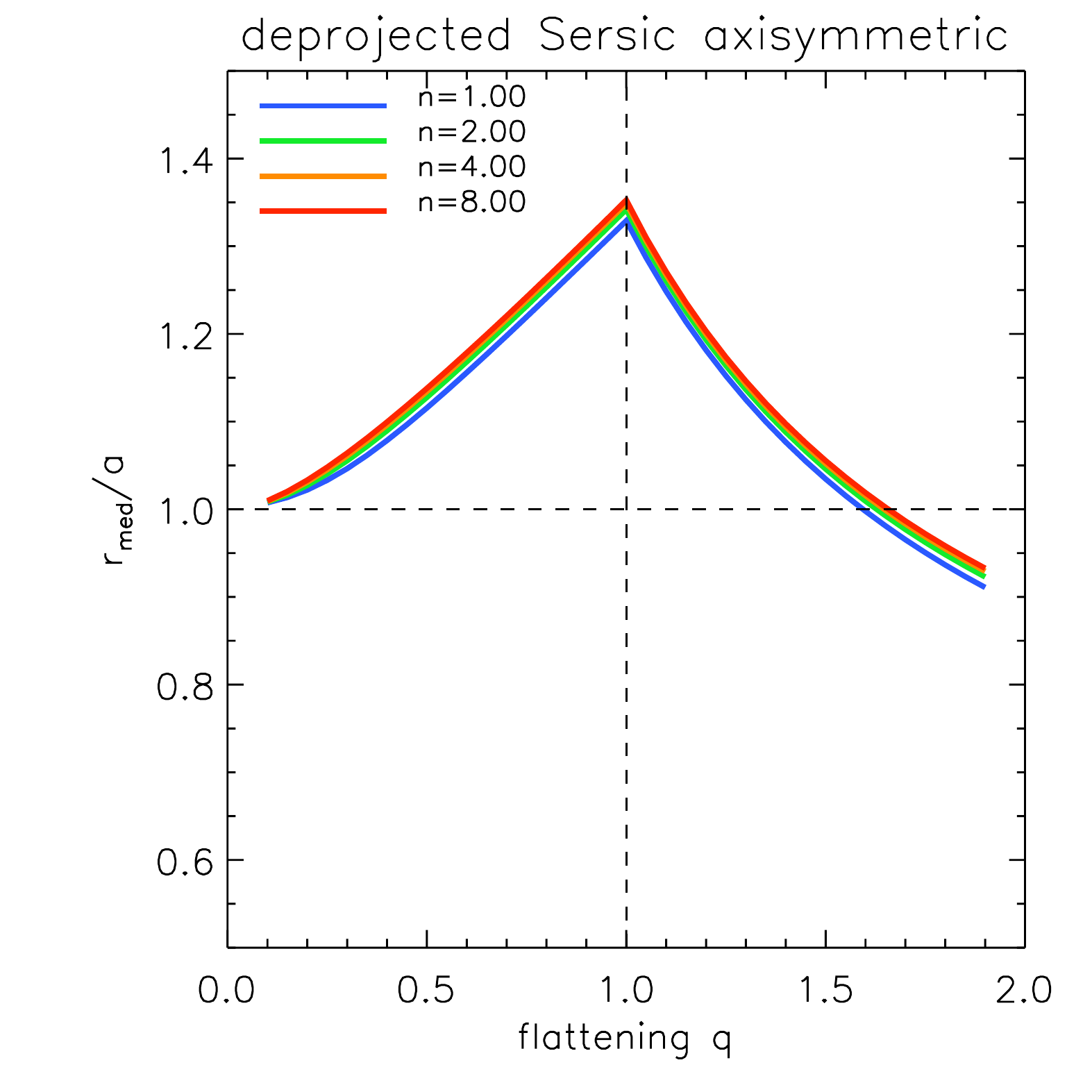}
 \includegraphics[width=0.48\textwidth]{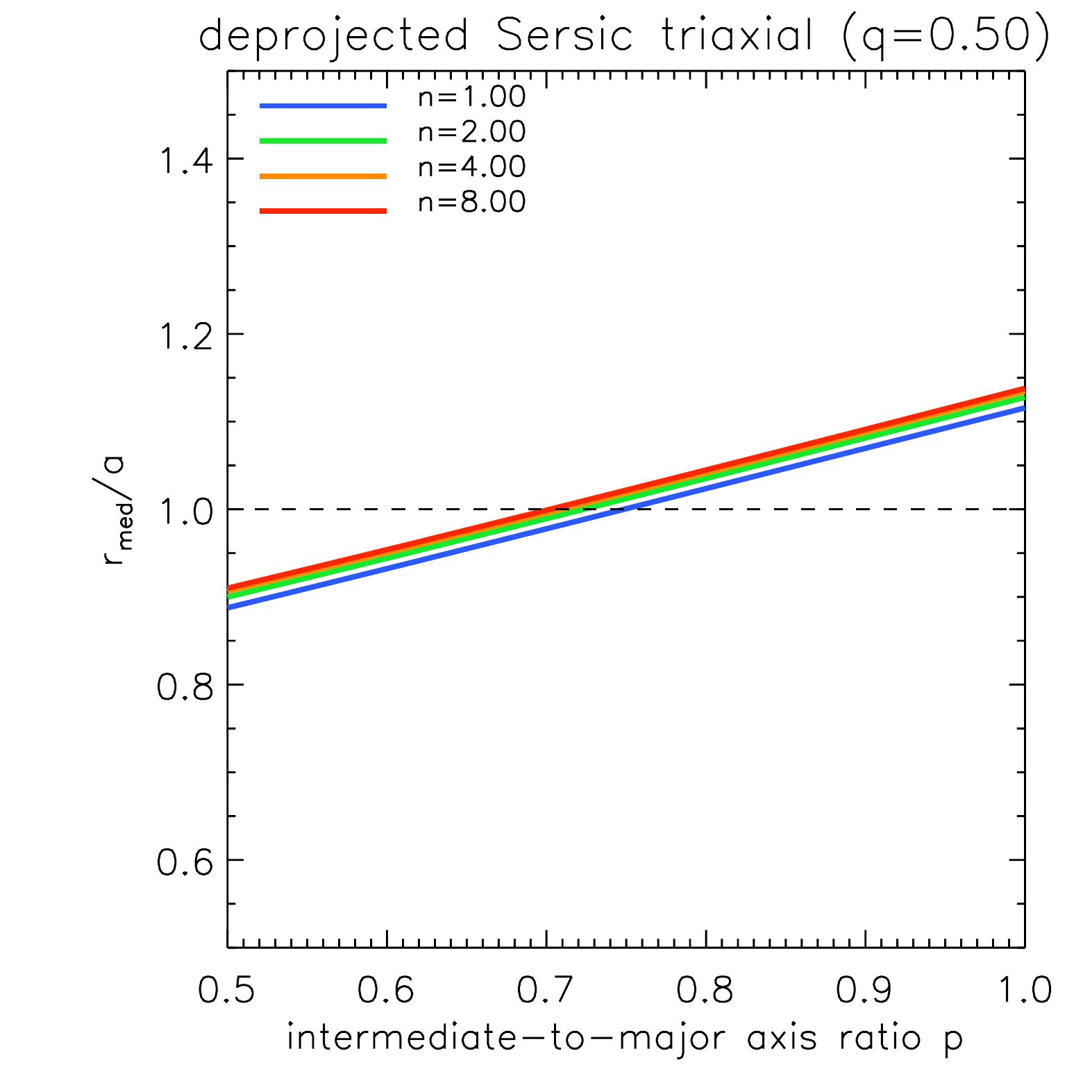}
  \caption{\small The median radius $\rmed$ of a de-projected S\'ersic profile normalized by the scale length $a$ when the enclosed mass is equal to half of the total mass (i.e., the intersections of the curves in Figure~\ref{fig:masssph} with the horizontal line at 0.5). The \emph{top} panel is for spheroidal shapes, either oblate axisymmetric ($q<1$) or prolate axisymmetrix ($q>1$), and the \emph{bottom} panel for fully triaxial shapes. In all cases, $\rmed/a$ depends on the intrinsic shape, but only mildly on the S\'ersic index $n$.} 
 \label{fig:rmed_intr}
\end{center}
\end{figure}

\section{Deprojected S\'ersic density profiles}
\label{S:sersicdens}

It has long been known that the stellar surface density profiles of
early-type galaxies and of spiral galaxy bulges are well described by a
\citet{sersic68} profile $\Sigma(R) \propto \exp[-(R/R_e)^{1/n}]$, with
the effective radius $R_e$ enclosing half of the total stellar mass.  A
key conceptual difference from cusped models is that the profile
does not converge to a particular inner slope on small scales.  
Unfortunately, the deprojection of the Sersic surface density profile cannot be expressed in common functions \citep[for special functions see][]{baes11}, but is well approximated by the analytic density profile of \citet{prugniel97}
\begin{equation}
  \label{eq:rhosersic}
  \rho(m) =  \frac{\rho_0}{m^{p_n}} \; \exp\left[-b_n\,m^{1/n} \right],
\end{equation}
where the inner negative slope is given by 
\begin{equation}
  \label{eq:psersic}
  p_n = 1 - \frac{0.6097}{n} + \frac{0.05563}{n^2}.
\end{equation}
The enclosed mass for the Prugniel-Simien model is
\begin{equation}
  \label{eq:masssersic}
  M(<m) = 4 \pi a b c \rho_0 \; 
  n \, b_n^{(p_n-3)n} \, \gamma[(3-p_n)n,b_n\,m^{1/n}],
\end{equation}
where $\gamma[p;x]$ is the incomplete gamma function, which in the
case of the total luminosity reduces to the complete gamma function
$\Gamma[p] = \gamma[p;\infty]$.

The expression for the surface density is, to high accuracy, the
S\'ersic profile
\begin{equation}
  \label{eq:surfsersic}
  \Sigma(m') = \Sigma_0 \;
  \exp\left[-b_n\,(m')^{1/n} \right].
\end{equation}
Given the enclosed projected mass
\begin{equation}
  \label{eq:2dmasssersic}
  M'(<m') = 2 \pi a' b' \Sigma_0 \; 
  n \, b_n^{-2n} \, \gamma[2n,b_n\,(m')^{1/n}],
\end{equation}
the requirement that the total intrinsic and projected mass have to be
equal yields a normalisation 
\begin{equation}
  \label{eq:normalizationsersic}
  \Sigma_0 = \frac{abc}{a'b'} \, \rho_0 \, 
  \frac{2\,\Gamma[(3-p_n)n]}
  {b_n^{(1-p_n)n}\,\Gamma[2n]}.
\end{equation}
The value of $b_n$ depends on the index $n$ and the choice for the
scale length. The latter is commonly chosen to be the effective radius
$R_e$ in the stellar surface density profile, which contains half of the
total stellar mass. We adopt a similar convention requiring that the ellipse
$m'=1$ contains half of the projected mass. This choice results in the
relation $\Gamma[2n] = 2\,\gamma[2n,b_n]$, which to high precision can
be approximated by \citep[cf.][]{ciotti99}
\begin{equation}
  \label{eq:approxb}
  b_n = 2\,n 
  - \frac{1}{3} 
  + \frac{4}{405} \, \frac{1}{n} 
  + \frac{46}{25515} \, \frac{1}{n^2}. 
\end{equation}
%

\section{Median radius}
\label{S:medianradius}

\begin{figure*}[th!]
\begin{center}
 \includegraphics[width=0.48\textwidth]{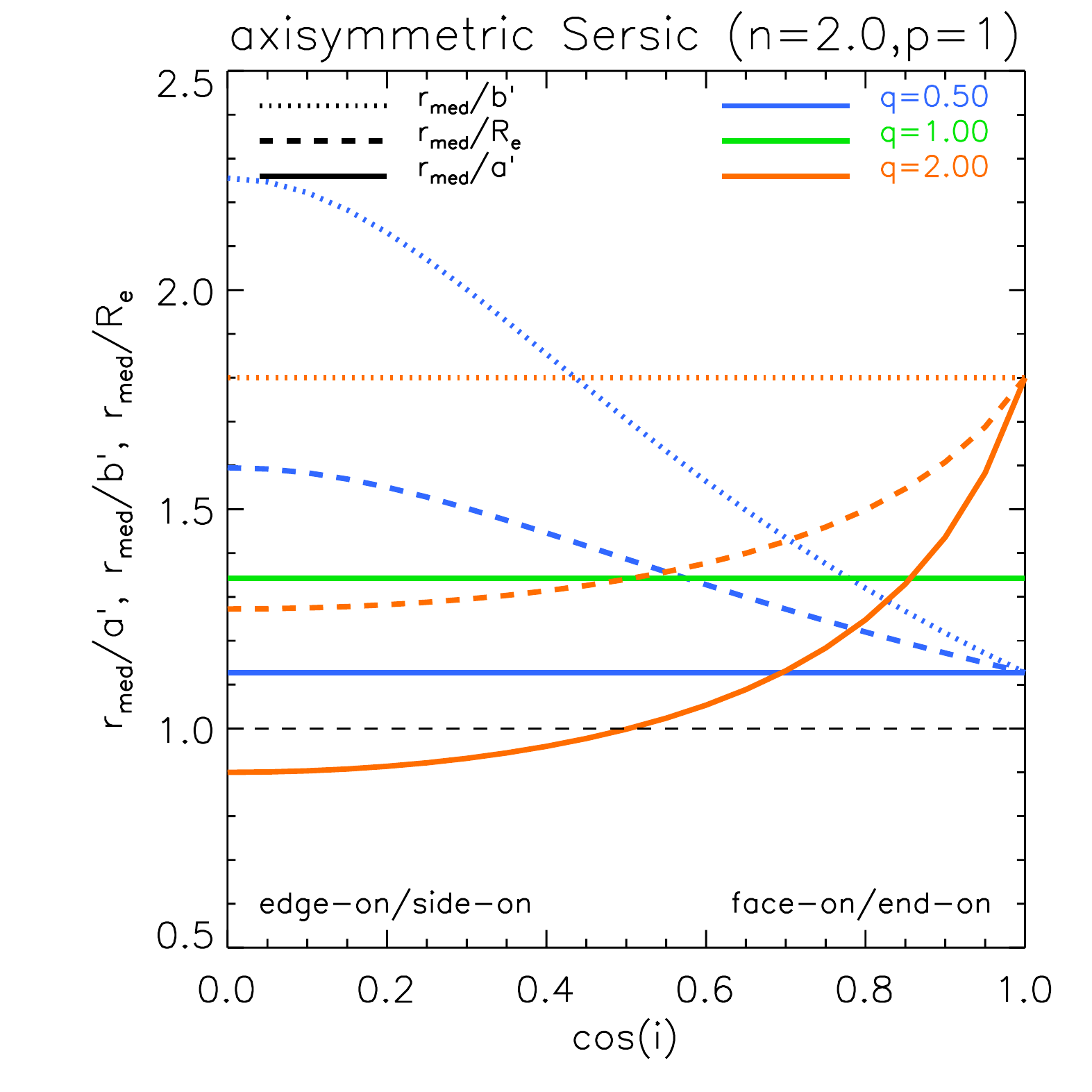}
 \includegraphics[width=0.48\textwidth]{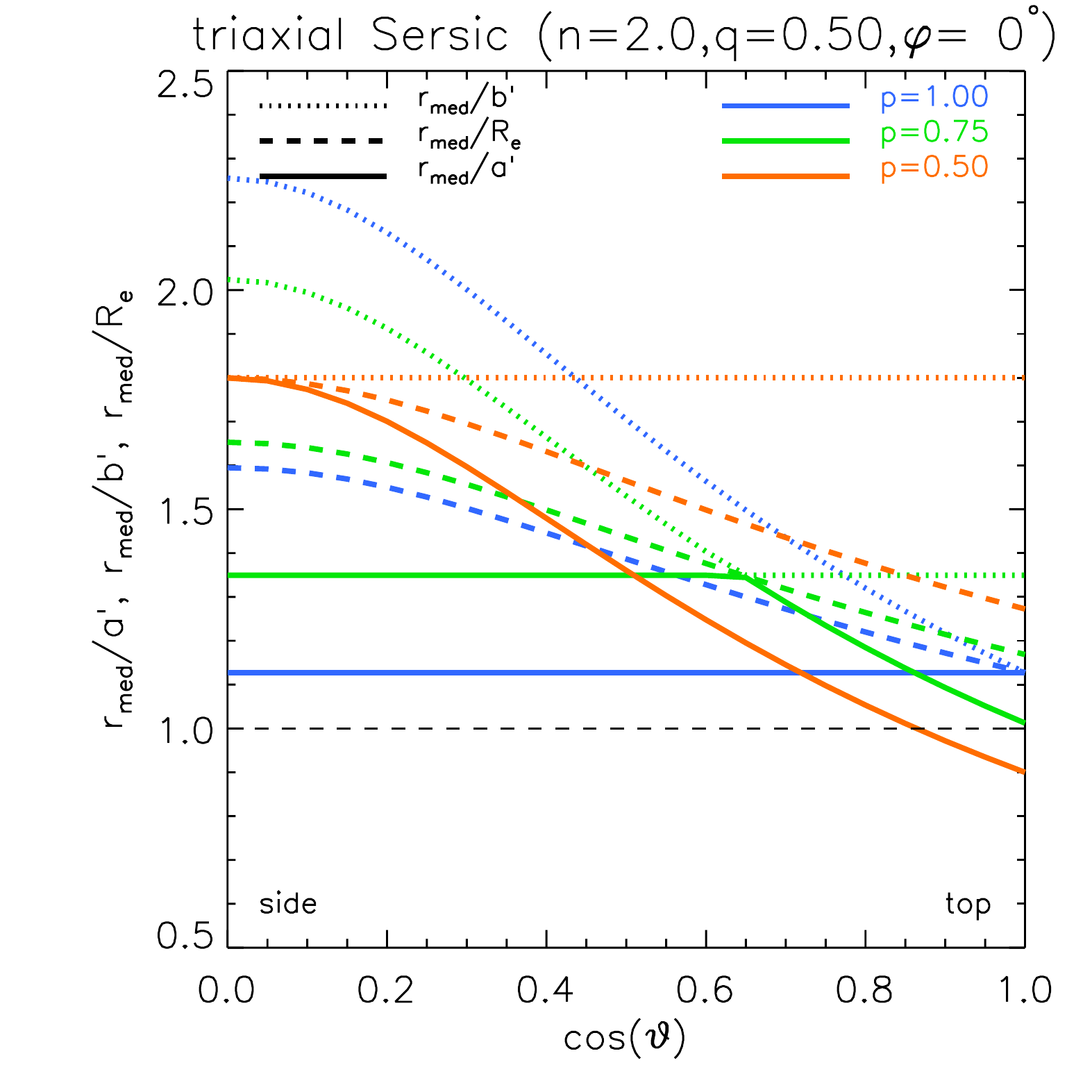}
 \includegraphics[width=0.48\textwidth]{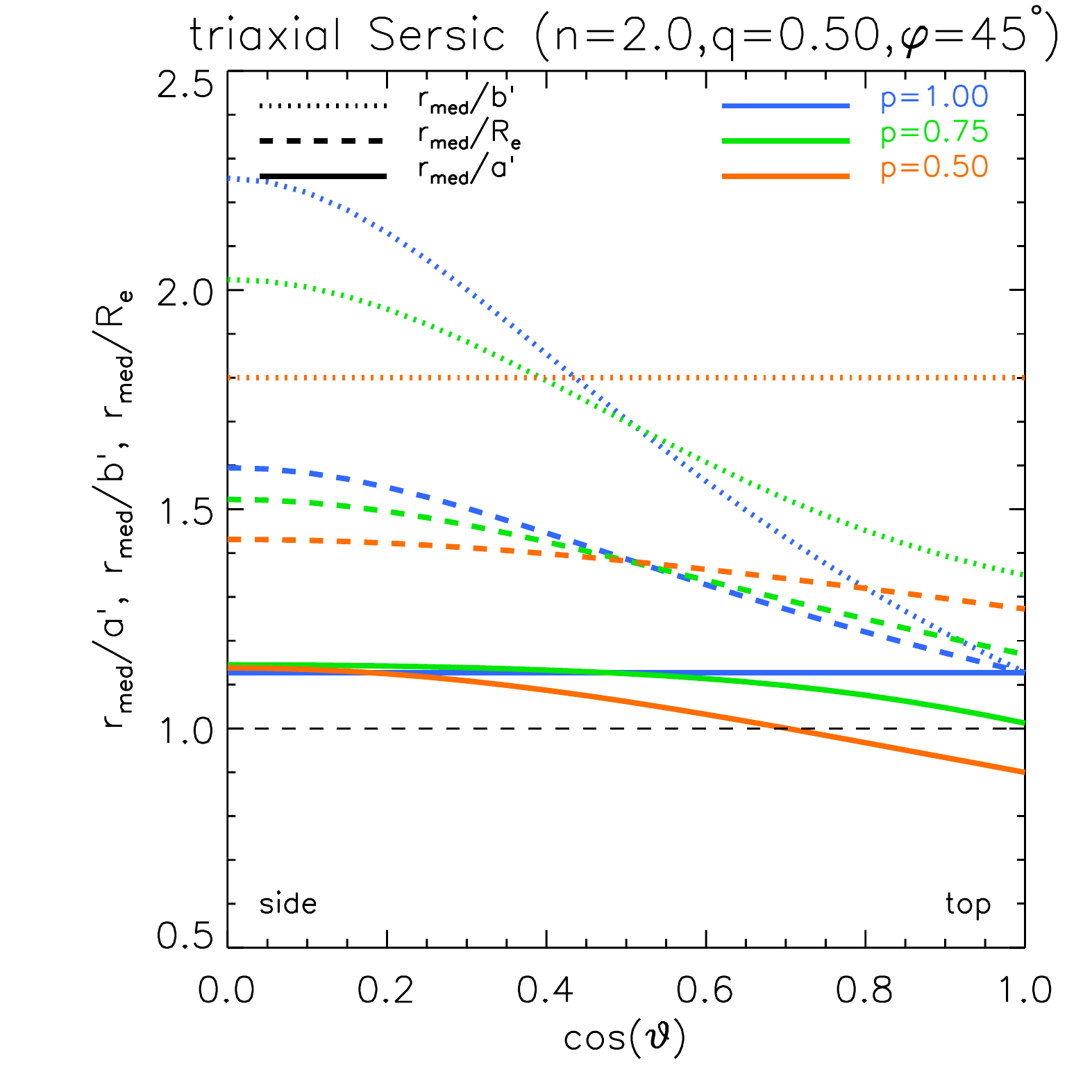}
 \includegraphics[width=0.48\textwidth]{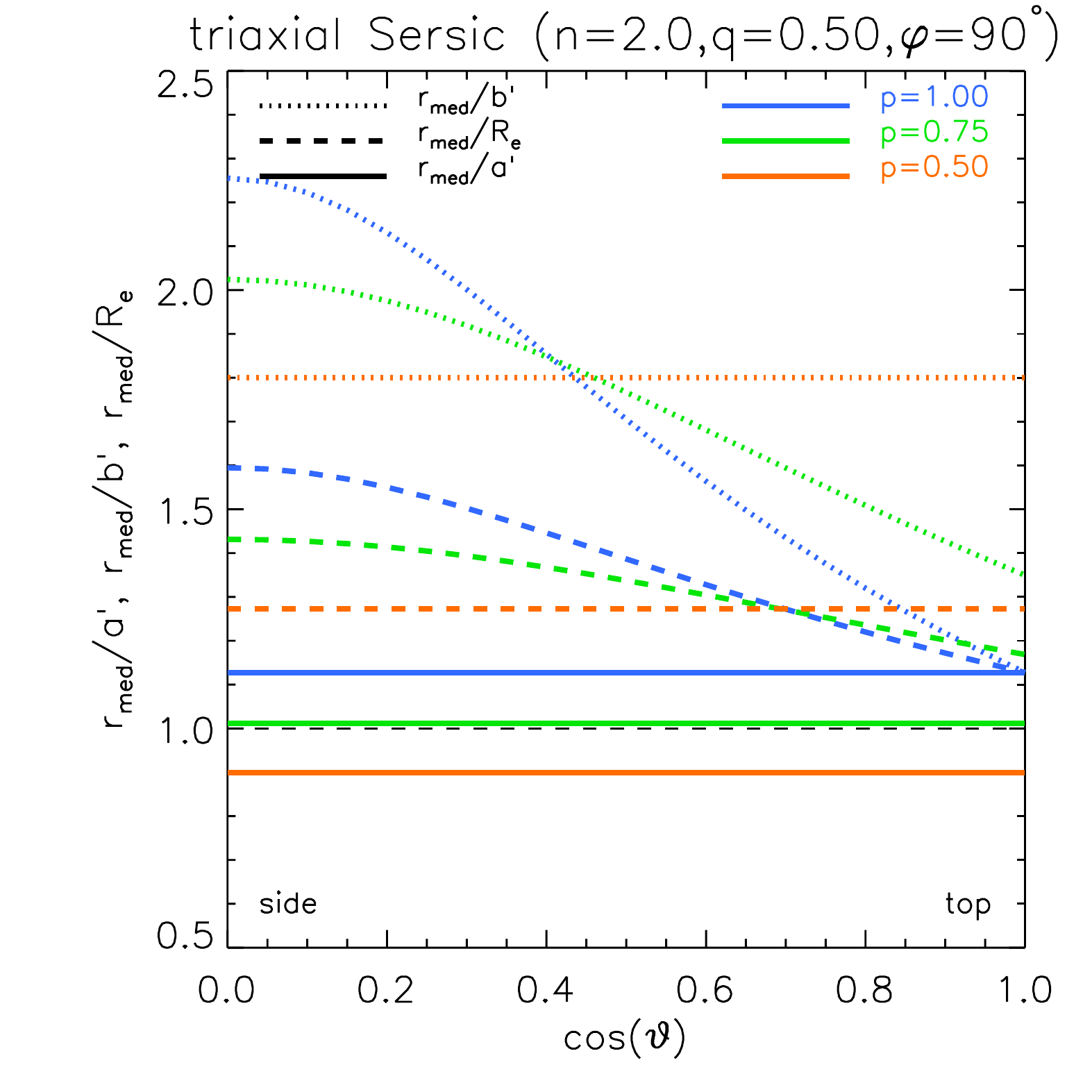}
  \caption{\small The median radius $\rmed$ of a deprojected S\'ersic profile normalized by the projected semi-major axis length $a'$ (solid curves) and semi-minor axis length $b'$ (dotted curves), as well as effective radius $R_e$ (dashed curves), as function of the polar viewing angle $\vartheta$. The \emph{top-left} panel is for the axisymmetric case with inclination $i$ ranging from $\cos i = 0$ for edge-on (side-on) to $\cos i = 1$ for face-on (end-on) in the oblate (prolate) axisymmetric case, with intrinsic flattening $q<1$ ($q>1$) indicated by the different colors. The \emph{other three panels} are for the triaxial case, which in addition to the polar viewing angle $\vartheta$, also depends on the azimuthal viewing angle $\varphi$ with values indicated above each panel. The intrinsic short-to-major axis ratio is fixed to $q=0.5$, whereas the colors indicate different intrinsic intermediate-to-major axis ratios $p$.} 
 \label{fig:rmed_proj}
\end{center}
\end{figure*}

We suggest as a robust measure for the size of a galaxy the median spherical radius. To infer this value numerically we have to find the spherical radius which encloses half of the stellar mass. The mass enclosed within a spherical radius $r$ follows from
\begin{equation}
  \label{eq:masssph}
  M(<r) = \int_0^r \tilde{r}^2 \rmd \tilde{r} \int_0^\pi \sin\theta \rmd \theta \int_0^{2\pi} \rmd \phi \; \rho[m(r,\theta,\phi)],
\end{equation}
with ellipsoidal radius $m(r,\theta,\phi) = (r/a) \, C_{pq}(\theta,\phi)$, where we have defined
\begin{equation}
  \label{eq:defCpqthph}
  C_{pq}^2(\theta,\phi) = \sin^2\theta \left( \cos^2\phi + \frac{\sin^2\phi}{p^2} \right) + \frac{\cos^2\theta}{q^2},
\end{equation}
and as before $p=b/a$ and $q=c/a$. Substituting the deprojected S\'ersic density of equation~\eqref{eq:rhosersic}, we see that the integral over radius is an incomplete gamma function, so that
\begin{eqnarray}
  \label{eq:sersicmasssph}
  M(<r) & = & 4 \pi a^3 \rho_0 \; n \, b_n^{(p_n-3)n} \int_0^1 \rmd \cos\theta \int_0^1 \rmd (\phi/2\pi)
  \nonumber\\
  && \times \gamma[(3-p_n)n,b_n\,(C_{pq} \, r/a)^{1/n}] \; C_{pq}^{-3}.
\end{eqnarray}
With the total stellar mass given by equation~\eqref{eq:masssersic} for $m\to\infty$, the median radius normalized by the scale length $\rmed/a$, follows upon numerically solving 
\begin{eqnarray}
  2\, p\, q\, && \gamma[(3-p_n)n] = \int_0^1 \rmd \cos\theta \int_0^1 \rmd (\phi/2\pi) 
  \nonumber\\ 
  && \times \gamma[(3-p_n)n,b_n\,(C_{pq} \, \rmed/a)^{1/n}] \; C_{pq}^{-3}.
  \label{eq:sersicrmed}
\end{eqnarray}
When $p=1$ in oblate axisymmetry the integral over $\phi$ can be discarded. In case of prolate axisymmetry, it is easiest to exchange $a$ and $c$, so that again $p=1$ and the integral over $\phi$ can be discarded, whereas $q>1$ and the resulting $\rmed/c$ has to be divided by $q$ to obtain  $\rmed/a$.

As can be seen from Figures~\ref{fig:masssph}~and~\ref{fig:rmed_intr}, the resulting $\rmed/a$ depends on the intrinsic axis ratios $p$ and $q$, but very little on S\'ersic index $n$. This implies that the median radius is robust against measurement errors in $n$.

In practice we cannot measure the intrinsic scale length $a$, but instead we measure the projected semi-major axis length $a'$ and semi-minor axis length $b'$. Using the relations~\eqref{eq:apbpellipse}, we infer the ratio of the median radius to these projected length scales. Figure~\ref{fig:rmed_proj} shows the resulting $\rmed/a'$ (solid curves), $\rmed/b'$ (dotted curves), and $\rmed/R_e$ (dashed curves) with the adopted definition $R_e^2=a'b'$, as function of the polar viewing angle $\vartheta$.\looseness=-1

In the axisymmetric case (top-left panel), the latter polar angle is the inclination $i$, ranging from $\cos i = 0$ for edge-on (side-on) to $\cos i = 1$ for face-on (end-on) in the oblate (prolate) axisymmetric case, with intrinsic flattening $q<1$ ($q>1$) indicated by the different colors. 

In the triaxial case, the polar angle $\vartheta$ ranges from viewing the short axis from the side  ($\cos\vartheta=0$) to from the top  ($\cos\vartheta=1$). In addition, the projection depends on the azimuthal viewing angle $\varphi$, which ranges from viewing from along the long axis ($\varphi=0$\dgr, top-right panel), intermediate between long and short axis ($\varphi=45$\dgr, bottom-left panel) to along the intermediate axis ($\varphi=90$\dgr, bottom-right panel). The intrinsic short-to-major axis ratio is fixed to $q=0.5$, whereas the colors indicate different intrinsic intermediate-to-major axis ratios $p$.\looseness=-1

In all four panels the blue curves correspond to the same oblate axisymmetric case $(p=1,q=0.5)$, which is independent of the azimuthal angle $\varphi$. Even though the red curves also correspond to the prolate case with the same intrinsic flattening, the orientation in the top-left panel is such that its symmetry long-axis is along the $z$-axis whereas in the remaining panels it is along the $x$-axis. Finally, all curves are for a S\'ersic profile with index $n=2$, but we have seen before that there is little variation with $n$.

\section{Applications and Limitations}
\label{S:applications-and-limitations}

\subsection{Inclination Estimates for Individual Galaxies}
\label{SS:inclination-estimates} 

For individual galaxies a best-effort inclination estimate
makes use of the measured projected axis ratio $b'/a'$ as well as
prior statistical or specific (i.e., from kinematics) knowledge
about its intrinsic shape.

We first obtain the probability $f(\vartheta,\varphi|\eps)$ of
viewing the galaxy at angles $(\vartheta,\varphi)$ given its
observed ellipticity $\varepsilon \equiv 1 - b'/a'$. 
Then we combine the intrinsic shape and viewing angle distribution by drawing from $f(p,q)$ and $f(\vartheta,\varphi|\eps)$ and compute the corresponding distribution of ratios $\rmed/a'$. For this we need to perform a straightforward one-dimensional numerical integral for any given intrinsic (triaxial) shape distribution $f(p,q)$:
\begin{equation}
    f(\vartheta,\varphi|\varepsilon) = 
    \frac{\varepsilon(1-\varepsilon)(2-\varepsilon)}{\cphq \sin{\varphi}} 
    \int_{0}^{1} \\
    \frac{f(p,q)\alpha^4}{p~q \sqrt{D}}\rmd{\alpha^2},
\end{equation}
where $\alpha \equiv a'/a$ and 
\begin{eqnarray}
    D \; & \equiv & \; \varepsilon^2(2-\varepsilon)^2\alpha^4  \nonumber\\ 
        && - 4 \tan^2\varphi\cthq (1-\alpha^2)[1-\alpha^2(1-\varepsilon)^2].
\end{eqnarray}
Finally, we combine the distribution in $\rmed/a'$ with the observed
$a'$ including measurement uncertainties to derive $\rmed$, or more
precisely the distribution $f(\rmed|a',b')$.

We note that shape and
size are likely correlated, introducing another level of complexity:
in that case the assumption adopted by, e.g., \citet{chang13a} and \citet{van-der-wel14a} of random viewing angles -- independent of measured size or axis ratio -- no longer holds.  \citet{zhang19} examine the implications for the reconstruction of the intrinsic shapes of galaxy populations based on projected shape \emph{and} size distributions (building to a large extent on the work we present now in this paper). They find that for mixed populations of prolate-like and oblate-like systems it is indeed important to do this analysis jointly. 

\subsection{Size Estimates for Individual Galaxies} 
\label{SS:size-estimates} 

\begin{figure*}[th!]
\begin{center}
 \includegraphics[width=0.48\textwidth]{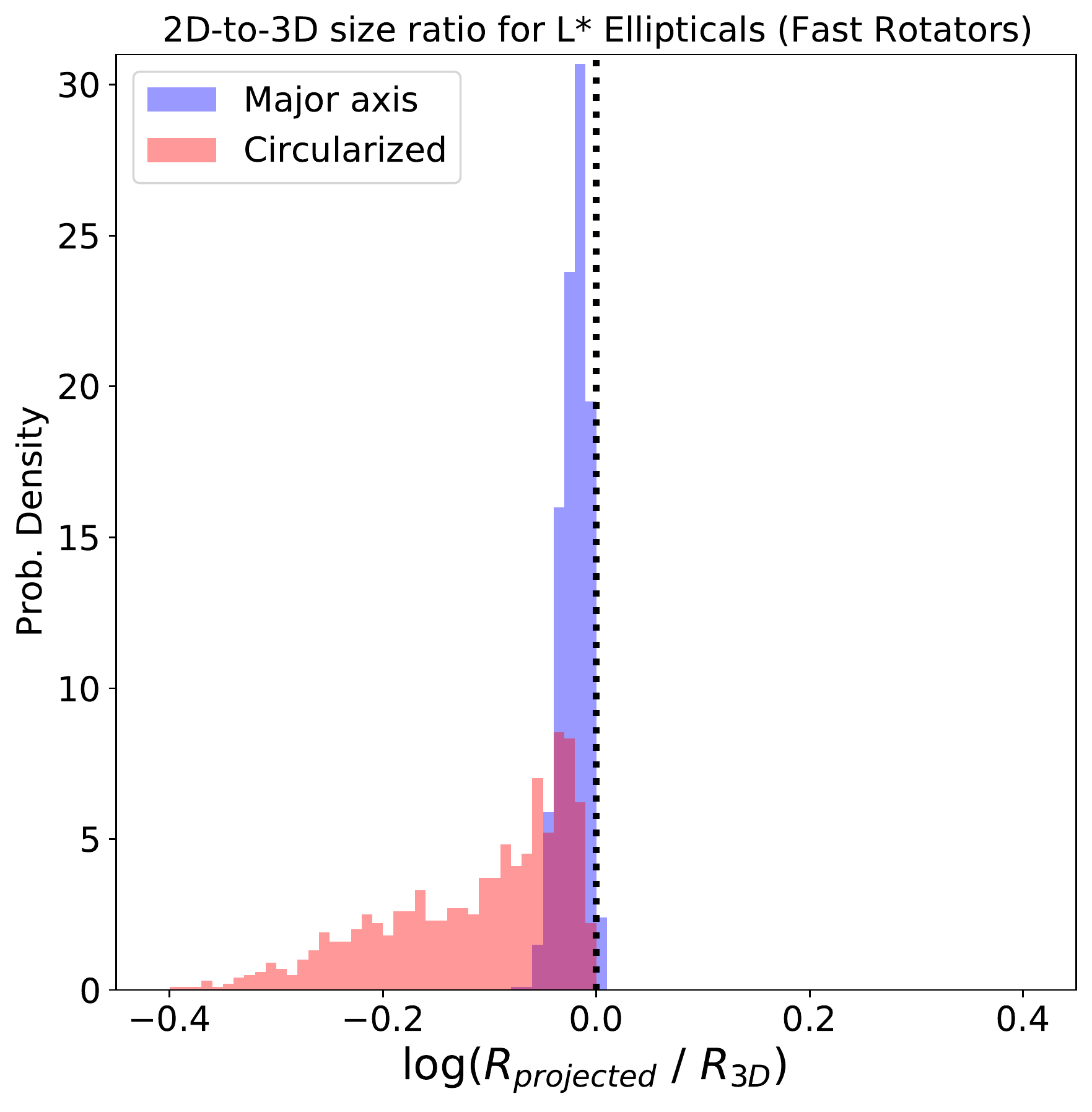}
 \includegraphics[width=0.48\textwidth]{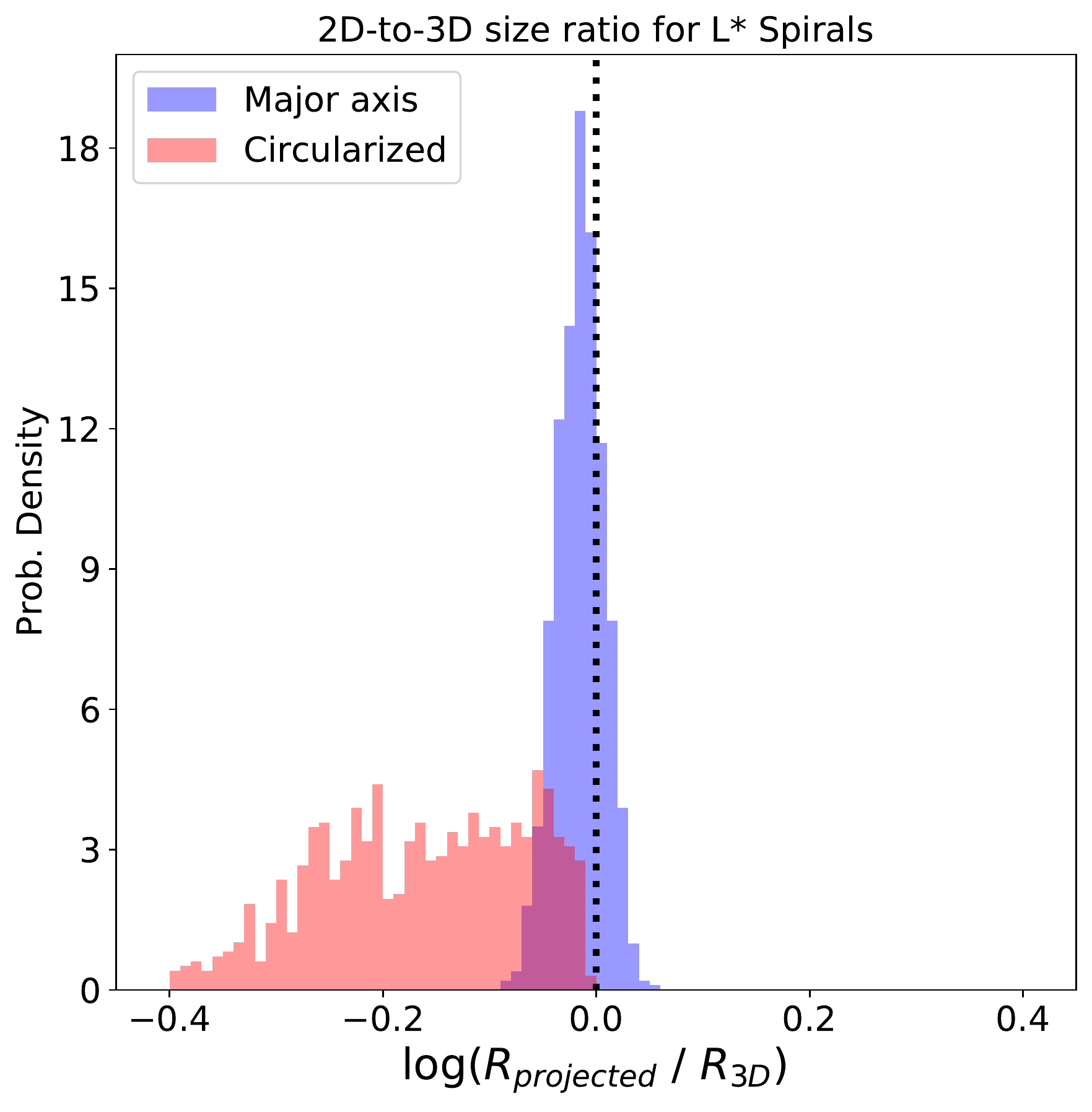}
 \includegraphics[width=0.48\textwidth]{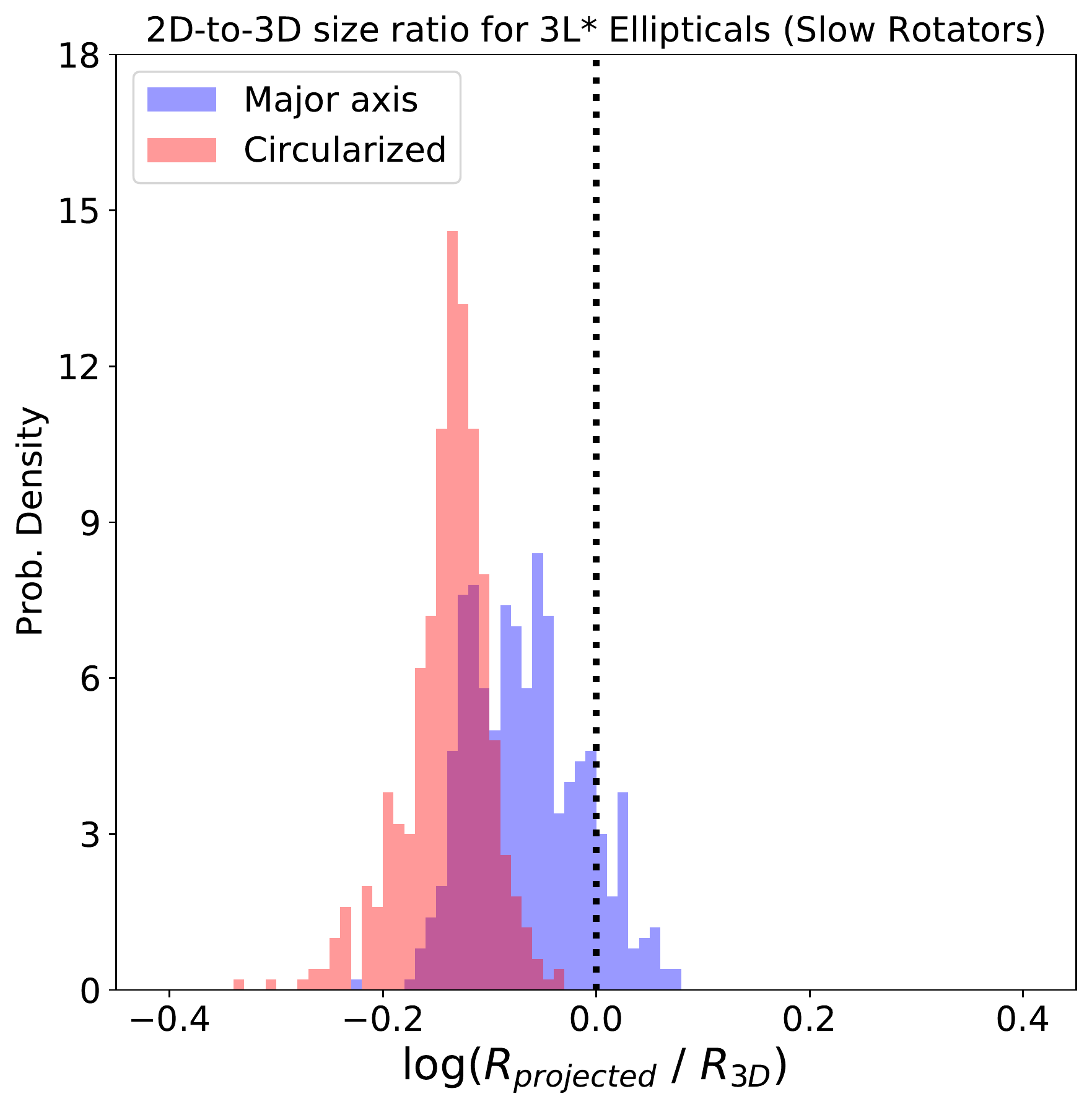}
 \includegraphics[width=0.48\textwidth]{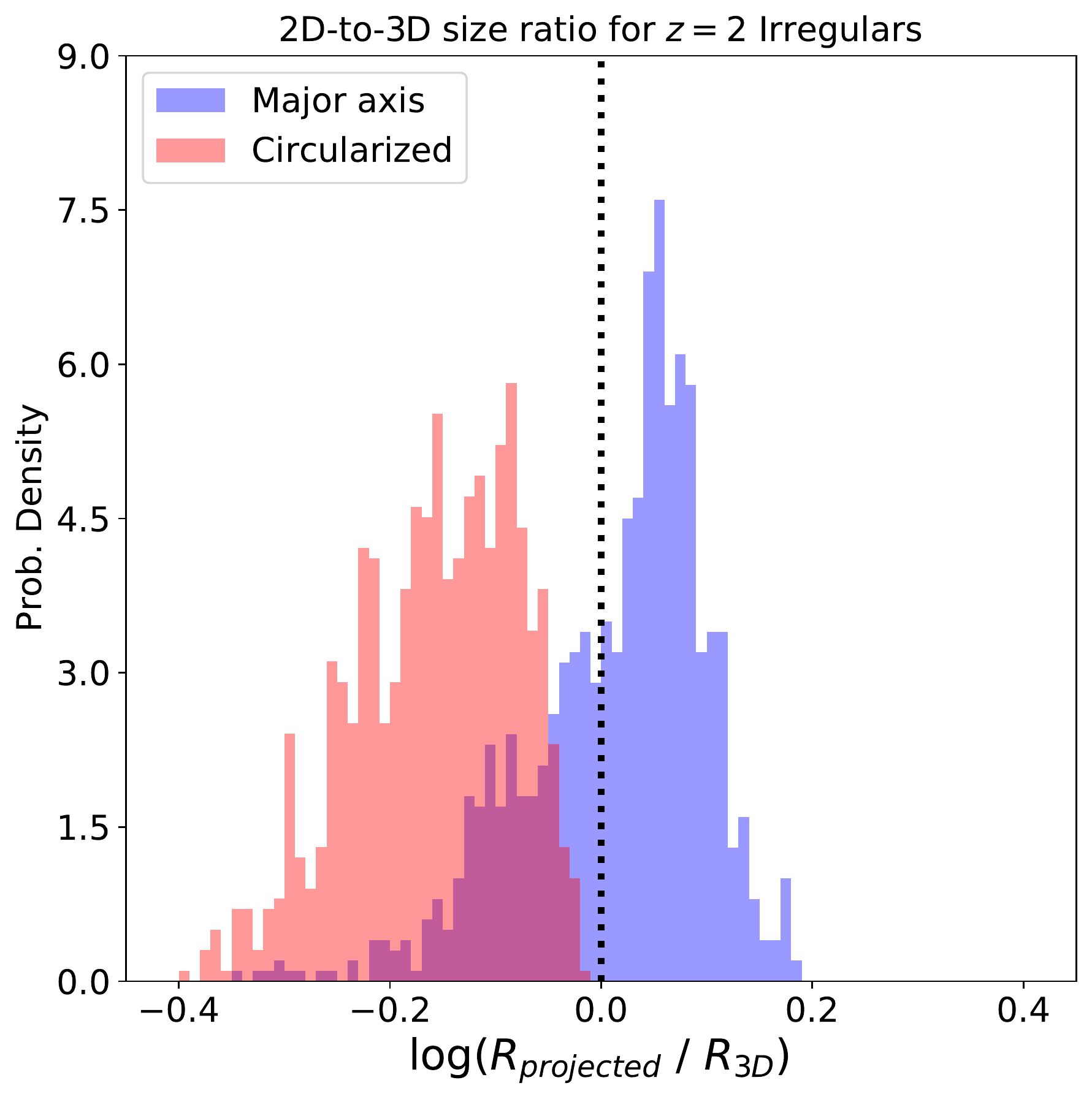}
  \caption{\small Simulated distributions (1000 objects each) of projected-2D to intrinsic-3D size ratios for four different galaxy types. The intrinsic shape distributions of the four types are given in Table \ref{tab:conversion-factor}. The blue histograms represent projected major axis sizes ($a'$) compared to intrinsic median sizes ($\rmed$), and red histograms show circularized sizes ($R_{e}$).} 
 \label{fig:hist}
\end{center}
\end{figure*}

A broad application of our work is to determine which (projected) size
measurement is the best proxy for the physically meaningful
$\rmed$. Traditionally, the circularized effective radius ($\sqrt{a'\,
  b'} = R_e$, in our notation) has been widely used, while others use
$a'$, the semi-major axis of the ellipse that encloses 50\% of the
total light.  Using our knowledge of the intrinsic shape distribution
of galaxies, we demonstrate here the differences between these approaches, and their 
sensitivity to instrinsic shape variations and viewing angle distributions.

\citet{chang13a} and \citet{van-der-wel14a} show that most galaxies in
the present-day universe have highly-flattened, nearly oblate geometries.  Crucially, this shape
distribution generally applies to different types of galaxies:
star-forming, quiescent, early-type, late-type, with the exception of
galaxies of very low and very high mass (see below).  We generate 
two simulated populations of such oblate and near-oblate objects seen at random viewing
angles. 

First, representing L$^*$ early-type galaxies (fast rotators), 
we assume a perfectly oblate population with intrinsic short-to-long axis ratios
$q=0.3\pm0.1$ \citep{chang13a}, where the error bar reflects the galaxy-to-galaxy Gaussian scatter.
The projected size distribution, relative to $\rmed$ is shown in Figure \ref{fig:hist} (top left). 
The major axis radius $a'$ reproduces $\rmed$ very closely, regardless of viewing angle, 
while $R_{e}$ strongly depends on viewing angle, inducing a large scatter. 
The median offset is just 18\%, but a factor of 2 difference is not a rare occurrence.

Second, representing L$^*$ late-type galaxies (spirals), we assume a near-oblate
shape distribution with $q=0.24\pm0.10$ and mild triaxiality $T=0.20\pm0.10$ 
to reflect the fact that most spirals are no perfectly circular when viewed face on.
Figure \ref{fig:hist} (top right) shows similar distributions as for L$^*$ early-type galaxies,
but with a much smaller median value for $R_{e}$: $R_{e}/\rmed=0.59$.
For the early types we see a slight offset from unity for $a'/\rmed$, which is partially
due to the lack of any triaxiality and partually due to a higher assumed S\'ersic index ($n=4$ compared to $n=1$).
The fraction of light projected into a cylinder compared to the light enclosed within a sphere is slightly 
smaller for high $n$ galaxies, but the difference here is just 2\%. Of course, for other apertures, 
e.g., enclosing 20\% or 80\% of the light these differences are much larger.


It is intuitively obvious that the measured semi-major axis of a disk
is invariant with inclination. Therefore, let us consider two classes
of galaxies with very different shapes. First, massive elliptical have been
shown to be strongly triaxial.  According to \citet{chang13a} massive
ellipticals have a triaxility of $T =  0.64\pm0.08$ and intrinsic short-to-long axis ratio of $q = 0.59\pm0.18$ \citep[see also][]{vincent05}. For a simulated population
of such objects seen at random viewing angles we find
medians of $a'/\rmed=0.85$ and $R_e/\rmed=0.73$ (see Figure \ref{fig:hist}, bottom left).
Perhaps surprisingly, even for massive ellipticals $a'$ provides a more accurate
estimate of $\rmed$ than $R_e$. At the same time, $R_e$ is more precise 
in the sense that the scatter in $R_e/\rmed$ is smaller.

Second, high-redshift, low-mass star-forming galaxies, among observed galaxy populations, deviate the most from oblate shapes and therefore provide the most
stringest test of our approach. Initiated by the discovery of so-called chain
galaxies \citep{cowie95}, we have learned that low-mass galaxies at
$z\gtrsim 2$ have very diverse geometries, ranging from oblate to
prolate \citep{ravindranath06, yuma12, law12, van-der-wel14a}.
Adopting $T=0.75\pm0.15$ and $q = 0.24\pm0.10$, we find
$a'/\rmed=1.08$ and $R_e/\rmed=0.71$ (see Figure \ref{fig:hist}, bottom right).
Once again, $a'$ provides a less biased galaxy size, and the scatter is similar for $R_{e}$ and $a'$

\begin{table}[t]\scriptsize
 \caption{Projected to intrinsic galaxy sizes for different  populations}
 \begin{tabular}{|c||c|c|c|c|}
\hline
 Population & $T$ & $q$ & $R_e/\rmed$ & $a'/\rmed$ \\
\hline
L$^*$ E        & 0             & $0.30 \pm 0.10$ & $0.82^{+14\%}_{-25\%}$ & $0.96^{+3\%}_{-3\%}$ \\
L$^*$ S        & $0.20\pm0.10$ & $0.24 \pm 0.10$ & $0.69^{+28\%}_{-23\%}$ & $0.97^{+5\%}_{-5\%}$ \\
3L$^*$ E       & $0.64\pm0.08$ & $0.59 \pm 0.18$ & $0.73^{+7\%}_{-9\%}$ & $0.85^{+15\%}_{-11\%}$ \\
$z=2$ Irr & $0.75\pm0.15$ & $0.24 \pm 0.10$ & $0.71^{+18\%}_{-19\%}$ & $1.08^{+13\%}_{-23\%}$ \\
\hline
\end{tabular}
  \tablecomments{Conversion factors from projected to intrinsic sizes for four galaxy populations: 
  L$^*$ early-type galaxies (Fast Rotators),
  L$^*$ spiral galaxies, 
  massive elliptical galaxies (Slow Rotators), 
  and low-mass irregular galaxies at redshift $z=2$.
  Each galaxy population has a different intrinsic shape distribution as indicated by triaxiality $T$ (see Eq.~\ref{eq:misalignment_psi}) 
  and short-to-long axis ratio $q$
  with error bars representing the intrinsic (Gaussian) scatter. 
  The median size ratios compare measured projected sizes (circularized $R_e$ or semi-major axis $a'$) to intrinsic size (median radius $\rmed$) with errors reflecting the 16th to 84th percentiles due to random variations in viewing angles and intrinsic shape.}
\label{tab:conversion-factor}
\end{table}

The overall conclusions we draw from this exercise are not as clear-cut as one would like. The decision to use $a'$ or $R_{e}$ depends on the situation. If the sample of relevance has a narrow range in
triaxiality $T$, then using $a'$ is preferable because of the lack of bias and small scatter with respect to $\rmed$. The range in oblateness does not affect this decision: the size distribution of a mix of very thick disks (even spheres) and thin disks is still much better described by $a'$ than $R_{e}$.
If, on the other hand, the sample spans a wide range of triaxialities -- or if the triaxiality is unknown -- then using $R_{e}$ is preferable because of the known, but relatively stable bias. The galaxy-to-galaxy scatter in size is larger in this case,
but this is unavoidable in the first place due to the lack of knowledge of the intrinsic shapes.



As a closing remark, let us stress that we explicitly assumed that
galaxies are transparent. The effects of viewing-angle dependent
extinction is likely the main uncertainty in determining the
light-weighted $r_{\rm{med}}$.  Furthermore, one would like to measure
mass-weighted sizes instead of light-weighted sizes. 
Mass-to-light ratio gradients due to age and metallicity gradients 
can have a large effect, with mass-weighted projected sizes that are typically 0.2 dex smaller than light-weighted sizes, to first order independent of galaxy type \citep[e.g.,][]{szomoru12, mosleh17, suess19}. Ideally, stellar surface mass density maps are used in combination with the methodology developed in this paper to arrive at size estimates that can directly be compared with theoretical models
and the results from numerical simulations.


Finally, the numerical implementation of the methods presented here can be made available upon request.


\acknowledgements{We thank the referee for helpful suggestions. This project has received funding from the European Research Council (ERC) under the European Union's Horizon 2020 research and innovation programme under grant agreement No 724857 (Consolidator Grant ArcheoDyn) and No 683184 (Consolidator Grant LEGA-C).}


\bibliographystyle{apj.bst}
\bibliography{mypapers.bib}

\begin{thebibliography}{}
\expandafter\ifx\csname natexlab\endcsname\relax\def\natexlab#1{#1}\fi

\bibitem[{{Baes} \& {van Hese}(2011)}]{baes11}
{Baes}, M., \& {van Hese}, E. 2011, \aap, 534, A69

\bibitem[{{Bezanson} {et~al.}(2009){Bezanson}, {van Dokkum}, {Tal},
  {Marchesini}, {Kriek}, {Franx}, \& {Coppi}}]{bezanson09}
{Bezanson}, R., {van Dokkum}, P.~G., {Tal}, T., {et~al.} 2009, \apj, 697, 1290

\bibitem[{{Chang} {et~al.}(2013){Chang}, {van der Wel}, {Rix}, {Holden},
  {Bell}, {McGrath}, {Wuyts}, {H{\"a}ussler}, {Barden}, {Faber}, {Mozena},
  {Ferguson}, {Guo}, {Galametz}, {Grogin}, {Kocevski}, {Koekemoer}, {Dekel},
  {Huang}, {Hathi}, \& {Donley}}]{chang13a}
{Chang}, Y.-Y., {van der Wel}, A., {Rix}, H.-W., {et~al.} 2013, \apj, 773, 149

\bibitem[{{Ciotti} \& {Bertin}(1999)}]{ciotti99}
{Ciotti}, L., \& {Bertin}, G. 1999, \aap, 352, 447

\bibitem[{{Cowie} {et~al.}(1995){Cowie}, {Hu}, \& {Songaila}}]{cowie95}
{Cowie}, L.~L., {Hu}, E.~M., \& {Songaila}, A. 1995, \aj, 110, 1576

\bibitem[{{de Zeeuw} \& {Franx}(1989)}]{de-zeeuw89}
{de Zeeuw}, T., \& {Franx}, M. 1989, \apj, 343, 617

\bibitem[{{Djorgovski} \& {Davis}(1987)}]{djorgovski87}
{Djorgovski}, S., \& {Davis}, M. 1987, \apj, 313, 59

\bibitem[{{Dressler} {et~al.}(1987){Dressler}, {Lynden-Bell}, {Burstein},
  {Davies}, {Faber}, {Terlevich}, \& {Wegner}}]{dressler87}
{Dressler}, A., {Lynden-Bell}, D., {Burstein}, D., {et~al.} 1987, \apj, 313, 42

\bibitem[{{Franx}(1988)}]{franx88}
{Franx}, M. 1988, \mnras, 231, 285

\bibitem[{{Genel} {et~al.}(2018){Genel}, {Nelson}, {Pillepich}, {Springel},
  {Pakmor}, {Weinberger}, {Hernquist}, {Naiman}, {Vogelsberger}, {Marinacci},
  \& {Torrey}}]{Genel2018}
{Genel}, S., {Nelson}, D., {Pillepich}, A., {et~al.} 2018, \mnras, 474, 3976

\bibitem[{{Kormendy}(1977)}]{kormendy77}
{Kormendy}, J. 1977, \apj, 218, 333

\bibitem[{{Law} {et~al.}(2012){Law}, {Steidel}, {Shapley}, {Nagy}, {Reddy}, \&
  {Erb}}]{law12}
{Law}, D.~R., {Steidel}, C.~C., {Shapley}, A.~E., {et~al.} 2012, \apj, 745, 85

\bibitem[{{Ludlow} {et~al.}(2019){Ludlow}, {Schaye}, {Schaller}, \&
  {Richings}}]{Ludlow2019}
{Ludlow}, A.~D., {Schaye}, J., {Schaller}, M., \& {Richings}, J. 2019, \mnras,
  488, L123

\bibitem[{{Mosleh} {et~al.}(2017){Mosleh}, {Tacchella}, {Renzini}, {Carollo},
  {Molaeinezhad}, {Onodera}, {Khosroshahi}, \& {Lilly}}]{mosleh17}
{Mosleh}, M., {Tacchella}, S., {Renzini}, A., {et~al.} 2017, \apj, 837, 2

\bibitem[{{Mowla} {et~al.}(2019){Mowla}, {van Dokkum}, {Brammer}, {Momcheva},
  {van der Wel}, {Whitaker}, {Nelson}, {Bezanson}, {Muzzin}, {Franx},
  {MacKenty}, {Leja}, {Kriek}, \& {Marchesini}}]{Mowla2019}
{Mowla}, L.~A., {van Dokkum}, P., {Brammer}, G.~B., {et~al.} 2019, \apj, 880,
  57

\bibitem[{{Navarro} \& {Steinmetz}(2000)}]{navarro00}
{Navarro}, J.~F., \& {Steinmetz}, M. 2000, \apj, 538, 477

\bibitem[{{Parsotan} {et~al.}(2021){Parsotan}, {Cochrane}, {Hayward},
  {Angl{\'e}s-Alc{\'a}zar}, {Feldmann}, {Faucher-Gigu{\`e}re}, {Wellons}, \&
  {Hopkins}}]{Parsotan2021}
{Parsotan}, T., {Cochrane}, R.~K., {Hayward}, C.~C., {et~al.} 2021, \mnras,
  501, 1591

\bibitem[{{Prugniel} \& {Simien}(1997)}]{prugniel97}
{Prugniel}, P., \& {Simien}, F. 1997, \aap, 321, 111

\bibitem[{{Ravindranath} {et~al.}(2006){Ravindranath}, {Giavalisco},
  {Ferguson}, {Conselice}, {Katz}, {Weinberg}, {Lotz}, {Dickinson}, {Fall},
  {Mobasher}, \& {Papovich}}]{ravindranath06}
{Ravindranath}, S., {Giavalisco}, M., {Ferguson}, H.~C., {et~al.} 2006, \apj,
  652, 963

\bibitem[{{Sersic}(1968)}]{sersic68}
{Sersic}, J.~L. 1968, {Atlas de galaxias australes}

\bibitem[{{Suess} {et~al.}(2019){Suess}, {Kriek}, {Price}, \&
  {Barro}}]{suess19}
{Suess}, K.~A., {Kriek}, M., {Price}, S.~H., \& {Barro}, G. 2019, \apjl, 885,
  L22

\bibitem[{{Szomoru} {et~al.}(2012){Szomoru}, {Franx}, \& {van
  Dokkum}}]{szomoru12}
{Szomoru}, D., {Franx}, M., \& {van Dokkum}, P.~G. 2012, \apj, 749, 121

\bibitem[{{Trujillo} {et~al.}(2004){Trujillo}, {Rudnick}, {Rix}, {Labb{\'e}},
  {Franx}, {Daddi}, {van Dokkum}, {F{\"o}rster Schreiber}, {Kuijken},
  {Moorwood}, {R{\"o}ttgering}, {van der Wel}, {van der Werf}, \& {van
  Starkenburg}}]{trujillo04}
{Trujillo}, I., {Rudnick}, G., {Rix}, H.-W., {et~al.} 2004, \apj, 604, 521

\bibitem[{{van der Wel} {et~al.}(2014{\natexlab{a}}){van der Wel}, {Franx},
  {van Dokkum}, {Skelton}, {Momcheva}, {Whitaker}, {Brammer}, {Bell}, {Rix},
  {Wuyts}, {Ferguson}, {Holden}, {Barro}, {Koekemoer}, {Chang}, {McGrath},
  {H{\"a}ussler}, {Dekel}, {Behroozi}, {Fumagalli}, {Leja}, {Lundgren},
  {Maseda}, {Nelson}, {Wake}, {Patel}, {Labb{\'e}}, {Faber}, {Grogin}, \&
  {Kocevski}}]{van-der-wel14}
{van der Wel}, A., {Franx}, M., {van Dokkum}, P.~G., {et~al.}
  2014{\natexlab{a}}, \apj, 788, 28

\bibitem[{{van der Wel} {et~al.}(2014{\natexlab{b}}){van der Wel}, {Chang},
  {Bell}, {Holden}, {Ferguson}, {Giavalisco}, {Rix}, {Skelton}, {Whitaker},
  {Momcheva}, {Brammer}, {Kassin}, {Martig}, {Dekel}, {Ceverino}, {Koo},
  {Mozena}, {van Dokkum}, {Franx}, {Faber}, \& {Primack}}]{van-der-wel14a}
{van der Wel}, A., {Chang}, Y.-Y., {Bell}, E.~F., {et~al.} 2014{\natexlab{b}},
  \apjl, 792, L6

\bibitem[{{Vincent} \& {Ryden}(2005)}]{vincent05}
{Vincent}, R.~A., \& {Ryden}, B.~S. 2005, \apj, 623, 137

\bibitem[{{Yuma} {et~al.}(2012){Yuma}, {Ohta}, \& {Yabe}}]{yuma12}
{Yuma}, S., {Ohta}, K., \& {Yabe}, K. 2012, \apj, 761, 19

\bibitem[{{Zhang} {et~al.}(2019){Zhang}, {Primack}, {Faber}, {Koo}, {Dekel},
  {Chen}, {Ceverino}, {Chang}, {Fang}, {Guo}, {Lin}, \& {Wel}}]{zhang19}
{Zhang}, H., {Primack}, J.~R., {Faber}, S.~M., {et~al.} 2019, \mnras, 484, 5170

\end{thebibliography}

\end{document}